\documentclass[12pt]{iopart}
\usepackage{iopams}
\usepackage{setstack}
\usepackage{cite}

\usepackage[breaklinks]{hyperref}
\hypersetup{colorlinks,urlcolor=black,citecolor=black,linkcolor=black,filecolor=black}

\usepackage[utf8]{inputenc}
\usepackage{graphicx}
\graphicspath{{figures/}}

\newcommand{\vibA}{\epsilon_b}
\newcommand{\mrvel}{{\bar\Omega}}
\newcommand{\rvel}{{\Omega}}
\newcommand{\T}{ \mathcal{T}}
\newcommand{\mean}[1]{\left<#1\right>}
\newcommand{\abs}[1]{\left|#1\right|}
\newcommand{\tilt}{\theta_{T}}
\newcommand{\sci}[1]{{\times}{10^{{#1}}}}
\newcommand{\eqref}[1]{\eref{#1}}
\newcommand{\scale}{0.7}

\begin{document}

\title{Sustained rotation in a vibrated disk with asymmetric supports}

\author{Gonzalo G.~Peraza-Mues$^{1,2}$ and Cristian F.~Moukarzel$^1$}

\address{$^1$ CINVESTAV del IPN, Appl.~Phys.~Dept.\\ 97310 M\'erida, Yucat\'an,
  M\'exico.}

\address{$^2$ Universidad Politécnica de Yucatán\\
  Carretera Mérida‐Tetiz. Km 4.5, Ucú, Yucatán, México}

\ead{ggperaza@gmail.com}

\vspace{10pt}
\begin{indented}
\item[]\today
\end{indented}

\begin{abstract}
  A single frictional elastic disk, supported against gravity by two
  others, rotates steadily when the supports are vibrated and the
  system is tilted with respect to gravity. Rotation is here studied
  using Molecular Dynamics Simulations, and a detailed analysis of the
  dynamics of the system is made. The origin of the observed
  rotational ratcheting is discussed by considering simplified
  situations analytically.  This shows that the sense of rotation is
  not fixed by the tilt but depends on the details of the excitation
  as well.
\end{abstract}

\vspace{2pc}

\submitto{\JSTAT}

\maketitle


\section{Introduction}\label{sec:intro}
Since the discussion by Feynman of the Feynman-Smoluchowski ratchet in his
famous lectures~\cite{feynman_feynman_2011}, much work has been devoted to the
study of ratcheting systems.  A system is said to ratchet if it is able to
rectify noise (e.g.\ thermal noise, random vibrations) into directed motion.
Examples in microscopic systems can be found in the field of molecular
motors~\cite{astumian_fluctuation_1994,reimann_brownian_2002,astumian_brownian_2002,hanggi_brownian_2005,hanggi_artificial_2009,reimann_introduction_2014,parrondo_energetics_2014},
while several granular systems have been
proposed~\cite{costantini_granular_2007,cleuren_granular_2007,van_den_broeck_microscopic_2004,eshuis_experimental_2010,gnoli_brownian_2013}
that display ratcheting on larger size scales. All ratcheting systems work out
of equilibrium, which allows them to escape the bounds imposed by the second law
of thermodynamics~\cite{feynman_feynman_2011,reimann_brownian_2002}.

Symmetry breaking of some sort is a requirement for
ratcheting~\cite{leff_maxwells_2014,hanggi_brownian_2005}. In granular
systems, an asymmetric intruder can be placed in a granular gas. The
asymmetries of the embedded object cause an imbalance of collisions
that makes this object either move
unidirectionally~\cite{costantini_granular_2007,cleuren_granular_2007},
or
rotate~\cite{van_den_broeck_microscopic_2004,eshuis_experimental_2010,gnoli_brownian_2013}. Simple
systems that rotate can also be
conceived~\cite{norden_ratchet_2001,altshuler_vibrot_2013}, in which a
chiral rotator is subjected to external excitation.

In this paper, a simple system is studied, that displays rotational
ratcheting: a single disk supported by two others against
gravity. This system is sketched in
Figure~\ref{fig:disks-notation}. When the support disks are vibrated,
numerical simulations and experiments~\cite{peraza-mues_rotation_2016}
show that the upper disk rotates steadily.  In our model system, the
rotating object is non-chiral, i.e.~reflection-symmetric. Reflection
symmetry is broken by tilting the system, which causes normal forces
at the contacts to differ. The aim of this work is to provide an
understanding of the microscopic origin of the rotational phenomenon.

Two rotational regimes can be identified, according to the intensity
of the vibration: a regime of gentle driving, where disks never lose
contact with each other, and a regime of medium driving, where disks
bounce against each other. Although persistent rotation is observed in
both dynamical regimes, it results from different dynamical processes
in each regime.  The case of rotation in the medium driving, bouncing
regime, has been already addressed numerically in previous
work~\cite{peraza-mues_rotation_2016}. The present investigation
focuses on the rotational phenomenon for low-intensity driving, that
is, in the regime of permanent contacts. In this regime, normal forces
between the upper disk and its contacts are never zero, but rotation
still happens because of the accumulation of frictional sliding. This
regime of sliding rotation is accessible to approximate methods of
analysis, which allow one to obtain a basic understanding of the
origin of the rotational imbalance. The results of our approximate
models can be satisfactorily tested against numerical simulations.

The rest of this paper is organized as follows. In
Section~\ref{sec:methods}, the 3-disk model, and the methods used in
numerical simulations are introduced. Numerical results showing
sustained rotation for random vibration of the supports are presented
in Section~\ref{sec:numer-simul}. In
Section~\ref{sec:disc-mech-rotat}, the origin of the rotational
imbalance is discussed, and two approximate models are solved for the
simpler case of deterministic periodic excitation of the
supports. Finally, our discussion and conclusions are presented in
Section~\ref{sec:conclusion}.
\section{Model and Methods}\label{sec:methods}
\begin{figure}[htbp]
  \centering
  \includegraphics[width=\scale\linewidth]{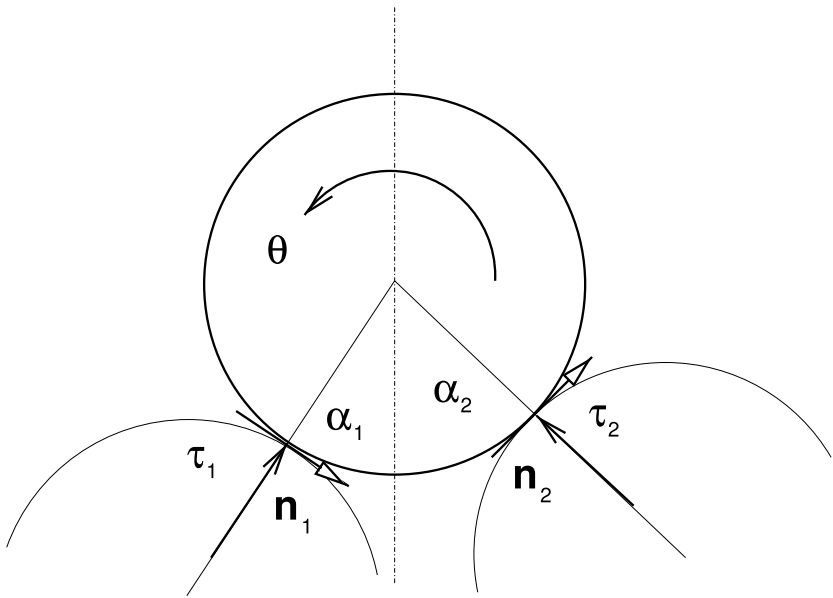}
  \caption{The 3-disk setup showing the direction of normal $n$ and tangential
    $\tau$ forces. $\theta$ is the angular coordinate of the upper disk. If
    angles $\alpha_1$ and $\alpha_2$ are not equal, the upper disk is found to
    rotate upon vibration of the supporting disks.\label{fig:disks-notation}}
\end{figure}

A setup of three disks of radius $R$ is considered, arranged as shown
in Figure~\ref{fig:disks-notation}. The freely-moving upper disk is
held against gravity by two support disks, whose excitatory motion is
externally prescribed, e.g.~as periodic or random vibration. Motion of
the supports causes fluctuations in the normal $n$ and tangential
$\tau$ forces at contacts 1 (left) and 2 (right). Under excitation,
and when $\alpha_1 \neq \alpha_2$, the upper disk is found to rotate
systematically in a given direction. The system behaves like a
ratchet, where the unbiased displacements of the supports are
rectified and the angular coordinate of the upper disk $\theta$ drifts
with mean rotational velocity $\mrvel$. This velocity $\mrvel$ depends
on several parameters, such as: the elastic properties of the disks,
the friction coefficient, the vibration intensity, the amount of tilt,
and the angle between contacts.

In this paper, the dependence of $\mrvel$ on the vibration intensity $\vibA$ and
tilt $\theta_T$ is explored by means of simulation and approximate
modelling. The tilt angle $\theta_T$ is defined as the angle between the
bisector of the contact lines joining the disk centers and gravity. It can be
calculated from the relation $\tilt=(\alpha_2 - \alpha_1)/2$, where $\alpha_1$
and $\alpha_2$ are the contact angles defined in
Figure~\ref{fig:disks-notation}. The angle between the contacts was chosen to be
$\alpha_1 + \alpha_2 = \pi/3$ (see Figure~\ref{fig:disks-notation}), as this is
the angle between contacts in the case of a two-dimensional close-packing of
equal disks.

For a given prescribed excitatory motion of the support disks, standard
molecular dynamics simulations were performed in order to follow the movements
of the upper disk. The equations of motions were integrated using a fifth-order
predictor-corrector algorithm~\cite{gear_automatic_1971} with a time-step
$dt=1\sci{-6}$. Each simulation ran up to $5\times 10^8$ time-steps for constant
conditions.

\subsection{Normal forces}\label{sec:low:nforces}
In our work, linear elasticity is assumed for the normal force between
two disks with centers at $\vec{r}_{a}$ and $\vec{r}_{b}$ with radii
$R_a$ and $R_b$. Defining \hbox{$\delta = (R_a+R_b)- |\vec r_b-\vec
  r_a|$}, one has
\begin{eqnarray}
  \label{eq:1}
  \vec{n} = \hat n_{ab}\left ( k_n \delta -
  \gamma_n u_n \right ) ,
\end{eqnarray}
where $\hat n_{ab}$ is the normal versor, $k_n$ is the compressive elastic
constant, $\gamma_n$ is a viscous constant, and $u_n$ is the normal relative
velocity (any sign) between disks. The viscous term $\gamma_n u_n$ accounts for
the energy dissipated through viscoelastic deformations of the disks. Notice that
(\ref{eq:1}) can become negative for two disks that are still ``in contact''
(the distance betewen their centers is smaller than the sum of their radii), if
they are moving appart from each other fast enough, because in this case the
contribution from the viscosity term is negative. Not correcting for this would
be unphysical, since, by definition, normal forces can only be compressive. In a
correct implementation of visco-elastic forces, one thus replaces (\ref{eq:1})
with
\begin{eqnarray}
  \label{eq:2}
  \vec{n} = \hat n_{ab} \ \rm{max}\left( k_n\delta -
  \gamma_n u_n,\ 0\right).
\end{eqnarray}
The physical meaning of this ``cutoff'' is easy to explain. When two
visco-elastic disks that are compressed toghether start to move apart, it takes
a certain time for them to expand and regain their original shape. Therefore, if
their (negative) relative velocity is large enough, they can become detached
from each other (their normal force becomes zero) even before the distance
between their centers becomes larger than the sum of their radii.

\subsection{Tangential forces}
A number of proposals have been put
forward~\cite{cundall_discrete_1979,shafer_force_1996,di_renzo_comparison_2004,kruggel-emden_review_2007,kruggel-emden_study_2008,kruggel-emden_applicable_2009}
to describe frictional forces between elastic bodies. The model for
tangential forces that is used in this work is a slightly modified
form of one originally proposed by Cundall and
Strack~\cite{cundall_discrete_1979,silbert_geometry_2002}. An
``elastic skin'' with tangential stiffness $k_t$ accounts for
tangential forces at each closed contact. The tangential force is
defined to be
\begin{eqnarray}
  \label{eq:3}
  \vec{\tau} =  -k_t \delta_t \hat t_{ab},
\end{eqnarray}
where $\hat t_{ab}$ is the tangential versor, and $\delta_t$ is the
accumulated tangential relative displacement between disks since they
last came in contact with each other.

The total tangential force is furthermore limited by the Amonton
condition
\begin{equation}
  \label{eq:36}
  \abs{\tau} \leq \mu n.
\end{equation}
If the Amonton condition is violated, $\delta_t$ is modified in order
to keep the total tangential force right at the frictional limit. This
adaptation represents the dissipative loss of elastic energy stored in
the ``elastic skin'', i.e., the particle's skin ``slips'' whenever
Amonton's limit is reached.

In the original model by Cundall and
Strack~\cite{cundall_discrete_1979}, $\delta_t$ is calculated as the
time integral of the relative tangential velocity of the surfaces in
contact. This method of calculating $\delta_t$ has been shown to cause
unrealistic deformation and ratcheting in granular
packings~\cite{mcnamara_microscopic_2008}, caused by an artificial
path dependence of the potential energy stored in the elastic
skin. Here an alternative, less error-prone, procedure was
implemented. This procedure allows one to calculate $\delta_t$
\emph{exactly}, directly from the knowledge of particle coordinates at
time $t$, plus one additional quantity that stores the memory of the
first contact and is modified upon sliding.

Let $\theta_{a}$ be the angular coordinate of a disk $a$. Upon general
rotations and displacements of their centers, the \textit{relative tangential
  displacement} $\lambda_{ab}$ of two disks $a$ and $b$ in contact is given by
\begin{equation}\label{eq:4}
  \lambda_{ab} =  R_{a} \theta_{a} +  R_{b} \theta_{b} -
  \beta_{ab} (R_{a} + R_{b}),
\end{equation}
where $\beta_{ab}$ is the angle made by the line that joins the centers of the
disks in contact with the $x$-axis. Notice that $\lambda$ is \emph{constant} for
two disks that roll on each other rigidly (without deformation of the skin) and
without slippage.

Assume that, when two disks are put in contact, their relative tangential
coordinate equals $\lambda^*$. If these disks are now moved slightly with
respect to each other, producing a change in $\lambda$ (\emph{without slip}),
tangential forces will develop. Tangential forces in our model were already
defined to depend linearly on the deformation of the
skin~\cite{cundall_discrete_1979}, that is:
\begin{equation}
  \tau =-k_t (\lambda -\lambda^*),
  \label{eq:tau}
\end{equation}
where it was assumed that no skin ``slippage'' has occurred as a
consequence of the tangential deformation. Therefore, upon stretching
of the skin, $\lambda^*$ still has the value that was defined at first
contact. When the tangential force is large enough to violate
Amonton's condition, the skin ``slides'' or ``slips''. This is
represented, in our implementation, by a change in $\lambda^{*}$ for
that contact, so as to maintain $|\tau|$ at its maximum possible
value, which is given by $\mu n$.

By way of example, assume that the Amonton condition is violated,
resulting in $-\mu n > k_t (\lambda -\lambda^*)$. In this case, one
redefines $\lambda^* = \lambda + \mu n/k_t$, such that the equality
$-\mu n = k_t (\lambda -\lambda^*)$ is restored. If, on the other
hand, the violation of the Amonton condition is such that $k_t
(\lambda -\lambda^*) > \mu n$, one redefines $\lambda^* = \lambda -
\mu n/k_t$ so as to have $k_t (\lambda -\lambda^*) = \mu n$. This
defines a sliding event, which simply amounts to a shift in
$\lambda^*$.

\section{Results for randomly vibrating supports}\label{sec:numer-simul}
While it is possible to numerically devise different displacement schemes for
the support disks, random vibration is of primary interest, as it demonstrates
the noise rectification properties of the system. In this section, numerical
results for such case are presented. Later, in
Sections~\ref{sec:synchr-rotat-supp} and~\ref{sec:asynchr-rotat-supp}, numerical
results for deterministic motion of the supports will be presented, along with
an analytic description of the dynamics.

For numerical simulations reported here, gravity was set to $g = 10$ and disks
were given a radius of $R=1$ and a mass $m = 0.1$. This results in a
gravitational force of $mg=1$. Units are arbitrary. Since our goal is to analyze
the rotational phenomenon and its causes, no attempt to relate the simulated
system to a physical one is made in this study.

Other physical constants can be related to the three quantities just introduced
as follows: With the given values for $g$, $R$, and $m$, the normal stiffness
$k_n$ can be defined from the ratio of the force required to compress a disk to
half is size and the force due to gravity, ${k_n R}/{mg}$. Since $R/mg = 1$,
this ratio is exactly equal to $k_n$. Similarly, $k_t$ is the ratio of the force
needed to obtain a tangential displacement of one radius and the force due to
gravity, ${k_t R}/{mg} = k_t$.

Numerical simulations were performed with a normal stiffness of
$k_n = 1\sci{3}$, tangential stiffness of $k_t = 1\sci{6}$, viscous damping
$\gamma = 10$, and friction coefficient $\mu=0.1$. Stiffness values where chosen
to make the disk tangentially stiff, while being relatively soft in the normal
direction. This allows us to explore a wider range of excitation amplitudes,
without the disk losing contact with the supports. Viscous dissipation is chosen
to keep normal oscillations in the underdamped regime, as it corresponds to a
damping ratio of $\gamma/2\sqrt{k_n m} = 0.5$.

It is also useful to define a characteristic time for the system. The time
needed for a disk to move a distance of one radius, starting from repose and
under the effect of gravity, is $t_g = \sqrt{2R/g} = 1/\sqrt{5}$. This way,
given a mean rotational velocity $\mrvel$, the quantity $\mrvel t_g$ is the
angle rotated during a time interval of $t_g$.

Random vibration is implemented by assuming that the coordinates $x$, $y$ and
$\theta$ of each support disk follow the dynamics of a white-noise-forced
harmonic oscillator. For example, for the $x$ coordinate, the following
stochastic equation of motion is integrated:
\begin{equation}
  \label{eq:3disklow:52}
  \ddot{x}(t) + 2c\omega \dot{x}(t) + \omega^2 x(t) = \xi(t),
\end{equation}
where $c$ is the damping ratio, $\omega$ is the natural frequency of
oscillation, and $\xi(t)$ is the forcing term. The random acceleration $\xi(t)$
has mean $\mean{\xi(t)} = 0$ and correlation
$\mean{\xi(t)\xi(t')} = 4c\omega^3{(\epsilon_bR)}^2\delta(t-t')$, where
$\epsilon_bR$ is the root mean square displacement of the vibration, and
$\epsilon_b$ is a dimensionless parameter controlling the amplitude of this
displacement with respect to the disk radius $R$. The motion of the support
disks is continuous and has a correlation time of
$t_c = 1/c\omega$~\cite{chandrasekhar_stochastic_1943}. The values for the
damping ratio and the natural frequency of the vibration were set to $c=0.5$ and
$\omega=20\pi$. The ratio of the correlation time $t_c$ to the characteristic
time $t_g$ is $t_c/t_g = \sqrt{5}/c\omega \approx 0.07$.

\begin{figure}
  \centering
  \includegraphics[width=0.8\linewidth]{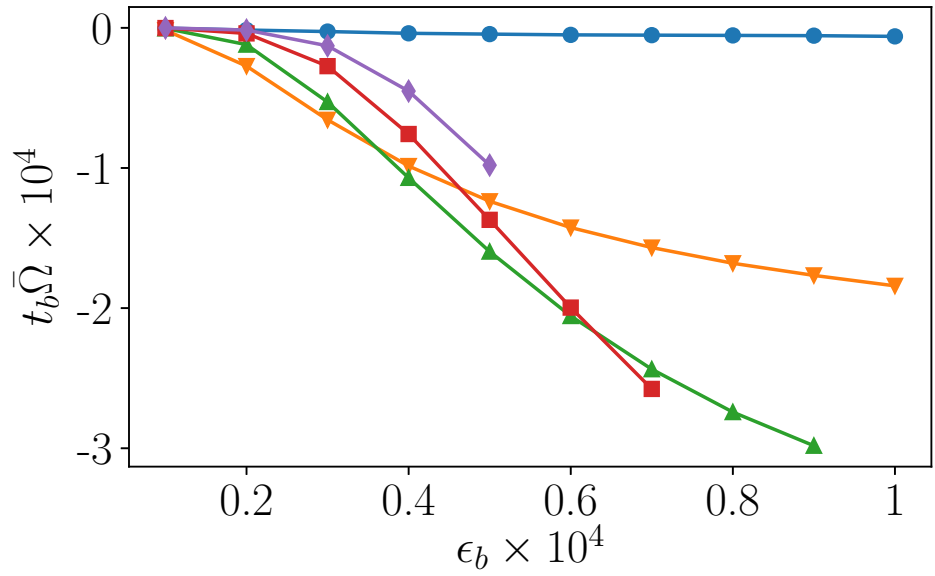}
  \caption{Numerical simulation results for the scaled mean rotational velocity
    $t_b\mrvel$ of the upper disk vs the scaled RMS amplitude $\epsilon_b$ of
    the vibrating support disks. Different lines correspond to different values
    of system tilt with respect to gravity, $\tilt=0\pi$ (circles),
    $\tilt=0.02\pi$ (triangles pointing down), $\tilt=0.04\pi$ (triangles
    pointing up), $\tilt=0.06\pi$ (squares), $\tilt=0.08\pi$
    (diamonds).}\label{fig:random-ex-vel}
\end{figure}

\begin{figure}
  \centering
  \includegraphics[width=0.8\linewidth]{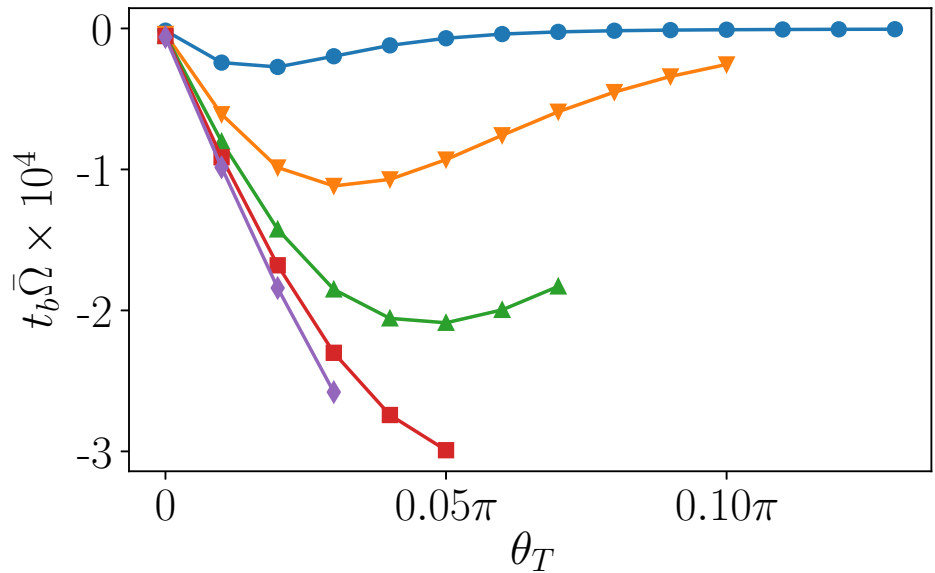}
  \caption{Numerical simulation results for the scaled mean rotational velocity
    $t_b\mrvel$ of the upper disk vs the system tilt $\tilt$. Different lines
    correspond to different values of amplitude $\vibA$ of the support
    vibrations, $\vibA=2\sci{-5}$ (circles), $\vibA=4\sci{-5}$ (triangles
    pointing down), $\vibA=6\sci{-5}$ (triangles pointing up), $\vibA=8\sci{-5}$
    (squares), $\vibA=1\sci{-4}$ (diamonds).}\label{fig:random-ex-tilt}
\end{figure}

Figure~\ref{fig:random-ex-vel} shows the scaled mean angular velocity
$t_g\mrvel$ of the upper disk vs $\epsilon_b$ for several values of tilt
$\tilt$. For $\tilt=0$, the disk does not rotate, as expected since the system
is reflection-symmetric in this case.
In Figure~\ref{fig:random-ex-tilt} the scaled angular velocity $t_b\mrvel$ is
shown vs the system tilt $\tilt$, for several values of the scaled amplitude
$\epsilon_b$. The rotational velocity is always clockwise (negative), and there
is a non-monotonic dependence of $\mrvel$ on $\tilt$.
In all simulations presented here, the upper disk never loses contact with the
supports. These results thus show that when the system is tilted, the breaking
of left-right symmetry allows the upper disk to rotate systematically in a given
direction. In the next section we explore in more detail how this asymmetry
induces rotation without assuming any particular displacement scheme for the
supports.

\section{Analysis of the dynamics of rotation}\label{sec:disc-mech-rotat}

In this Section, the microscopic mechanisms that cause rotation in the
3-disk system are explored.  Before starting such analysis, it is
useful to describe the relation between our 3-disk system and a
one-dimensional system subjected to similar frictional constraints,
that also exhibits drift under random excitation, namely the
elasto-plastic oscillator (EPO).

\begin{figure}
  \centering
  \includegraphics[width=\scale\linewidth]{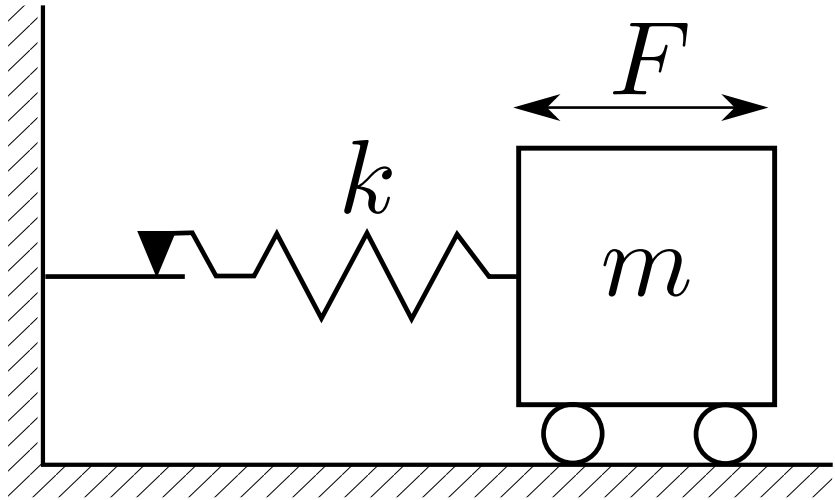}
  \caption{The undamped elasto-plastic oscillator. An externally excited mass
    $m$ coupled to a linear spring, of stiffness $k$, in series with a
    frictional slip-joint. The joint slides whenever magnitude of the force at
    the spring reaches a threshold value.}\label{fig:epo}
\end{figure}

\begin{figure}
  \centering
  \includegraphics[width=\scale\linewidth]{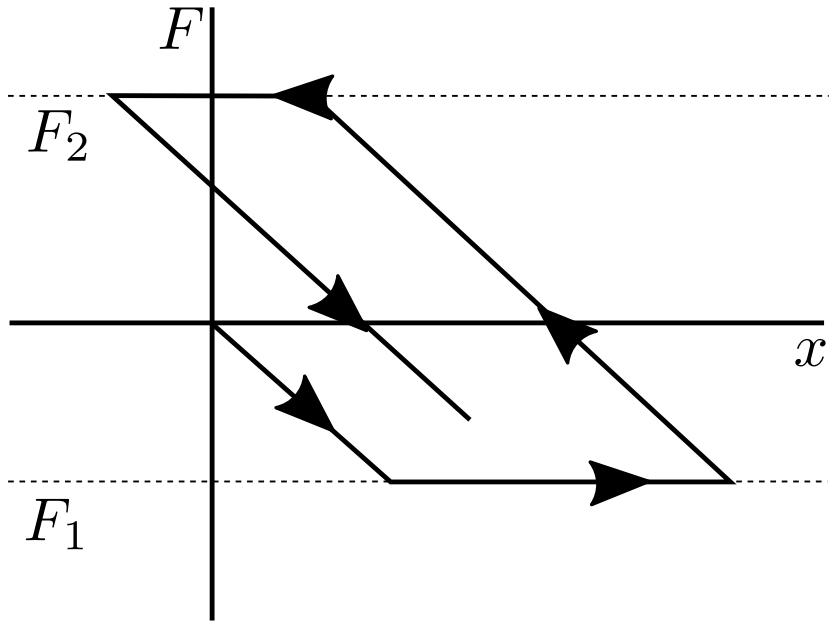}
  \caption{A possible evolution of the force on the spring in the EPO
    (Figure~\ref{fig:epo}). The force changes linearly with
    displacement as long as $F_1<F<F_2$. Once either limit is reached
    the force remains constant. If $F_1< F_2$, the lower limit is
    reached more easily, leading to a systematic displacement of the
    equilibrium point of the oscillator.}\label{fig:epo-force}
\end{figure}

A model for the undamped EPO is displayed in Figure~\ref{fig:epo}. It
consists of an externally excited mass $m$, coupled to a linear spring
in series with a frictional slip-joint. Due to the fictional
slip-joint, the force-displacement response of the EPO is
non-linear. Whenever the magnitude of the force on the spring reaches
a predefined threshold, the joints slide and the force remains
constant. The response of the EPO (and its generalizations) to both
harmonic
forcing~\cite{caughey_sinusoidal_1960,jennings_periodic_1964,iwan_steady-state_1965,masri_forced_1975,miller_gregory_r._periodic_1988,capecchi_periodic_1990,capecchi_periodic_1991,capecchi_asymptotic_1993,chatterjee_periodic_1996,liu_steady_2004,ahn_il-sang_dynamic_2006,csernak_periodic_2006,challamel_stability_2007-1,challamel_stability_2007}
and random
forcing~\cite{caughey_random_1960,karnopp_plastic_1966,iwan_response_1968,vanmarcke_probabilistic_1973,lutes_random_1974,grossmayer_elastic-plastic_1979,wen_equivalent_1980,ditlevsen_plastic_1993,bouc_ratcheting_1998,bouc_drifting_2002,feau_probabilistic_2008}
has been extensively studied before.
Figure~\ref{fig:epo-force} shows a possible force-displacement response for the
asymmetric EPO, starting at the equilibrium position at $x=0$. Under the effect
of the forcing, the mass starts to move in the positive direction, and the force
$F$ changes linearly, moving towards $F_1$. At the limit $F_1$, the slip-joint
starts sliding. As the mass continuous sliding, the force is kept constant at
$F_1$. Eventually, the direction of motion is reversed, and the joint stops
sliding. The force behaves linearly again and starts increasing towards
$F_2$. Upon reaching $F_2$, the joint slides again and the force remains
constant until motion once again reverses direction. Notice that, after sliding,
the equilibrium point has been displaced by an amount equal to the net sliding
distance, the force crossing zero at different values for $x$. When the forcing
cycle ends, the oscillator has experienced a positive drift.

Whenever the slip-joint slides, the equilibrium position moves in the
direction of sliding. If the frictional limits of the EPO are
symmetric, i.e., if $\abs{F_1} = \abs{F_2}$, sliding is equally likely
in either direction. The EPO experiences normal
diffusion~\cite{bouc_drifting_2002}, and its displacement averages to
zero. If, on the other hand, $\abs{F_1}\neq \abs{F_2}$, sliding will
be biased towards the smaller threshold. In this case, the mass will
drift systematically due to net sliding displacements in one direction
being larger than in the other. For example, if $F_1 < F_2$, the mass
slides more often to the right, making the average velocity of the EPO
positive. Such a drift has been studied
in~\cite{challamel_stability_2007} for a harmonically forced EPO, and
in~\cite{bouc_ratcheting_1998} for random forcing.

Each contact of the 3-disk system behaves similarly to an EPO.\@ Whenever the
relative tangential displacement of the upper disk and the support reaches the
limit imposed by the Amonton condition (equation~\eqref{eq:36}), the contact
slides. If sliding is biased, the upper disk rotates systematically upon being
excited. The system can be rationalized as two slip-joints, one at each contact,
which are rigidly coupled to each other. This coupling makes the 3-disk system
notably harder to analyze than the EPO.\@ Additionally, the Amonton limits
for different directions of sliding are not fixed, because normal forces at each
contact are allowed to change in response to the relative displacement of the
disks.

The forces acting on the disk are the normal and tangential forces at the
contacts, written as $n_i$ and $\tau_i$ for contact with the support disk $i$,
where $i=1,2$. Newton's second law applied to the upper disk's center of mass
results in the equations of motion:
\begin{eqnarray}
  \label{eq:13}
  m\ddot x = n_1\sin\alpha_1 - n_2\sin\alpha_2 + \tau_1\cos\alpha_1 + \tau_2\cos\alpha_2\\
  \label{eq:14}
  m\ddot y = n_1\cos\alpha_1 + n_2\cos\alpha_2 - \tau_1\sin\alpha_1 +
  \tau_2\sin\alpha_2 - mg,
\end{eqnarray}
where $x$ and $y$ are the coordinates of the upper disk's center, $m$ is the
upper disk's mass, $g$ is the gravitational acceleration, and the angles
$\alpha_1$ and $\alpha_2$ are defined in Figure~\ref{fig:disks-notation}.

Similarly, the total torque on the upper disk, caused by tangential
forces $\tau_1$ and $\tau_2$, is given by
\begin{equation}
  \label{eq:7}
  \T = R_0(\tau_1 + \tau_2),
\end{equation}
where $R_0$ is the upper disk's radius.

Normal and tangential forces depend on the relative distance between
the upper disk and the supports (see Section~\ref{sec:methods}),
making equations~\eqref{eq:13} trough~\eqref{eq:7} a set of coupled
differential equations. The system is non-linear because tangential
forces, which are subjected to the Amonton condition, introduce a
discontinuity into the equations each time a contact slides or ceases
sliding. This makes the equations of motion hard to solve, and
motivates the introduction of the following approximation to make the
system tractable.  In the regime of large support displacements and
low friction coefficient $\mu$, it is expected that contacts will
remain sliding most of the time. It is reasonable, then, to ignore the
duration of elastic deformations of the tangential skin and assume
that contacts are always sliding. From now on, this assumption will be
referred as the permanent-sliding approximation. The assumption is the
opposite of the one usually made in analytical treatments of the
EPO. For the EPO, it is often assumed~(see, for example,
references~\cite{karnopp_plastic_1966,bouc_ratcheting_1998,ditlevsen_plastic_1993})
that the slip-joint rarely slides, such that the dynamics of the
elastic regime is dominant. In the numerical simulations reported in
Section~\ref{sec:numer-simul}, it was verified that contacts slide
most of the time, in accordance with the proposed approximation. In
alternative simulations (not shown), where the supports move with low
intensity and sliding is rare, sustained rotation of the upper disk
was never observed numerically.

Under the assumption of permanent sliding, Amonton's equation~\eqref{eq:36}
becomes an equality, and can be used to reduce the number of unknowns by
rewriting equations~\eqref{eq:13} and~\eqref{eq:14} in a form that only involves
tangential forces,
\begin{eqnarray}
  \label{eq:37}
  \mu m\ddot x = \abs{\tau_1}\sin\alpha_1 - \abs{\tau_2}\sin\alpha_2
  + \mu\tau_1\cos\alpha_1 + \mu\tau_2\cos\alpha_2\\
  \label{eq:38}
  \mu m\ddot y = \abs{\tau_1}\cos\alpha_1 + \abs{\tau_2}\cos\alpha_2
  - \mu\tau_1\sin\alpha_1 + \mu\tau_2\sin\alpha_2 - mg.
\end{eqnarray}
The absolute values in equations~\eqref{eq:37} and~\eqref{eq:38} can be split
into four different cases, depending on the signs of $\tau_1$ and $\tau_2$. Each
of these four cases can be identified with a different sliding configuration of
contacts in the 3-disk system. These configurations are referred to as
$\{S^{++},S^{--}, S^{+-}, S^{-+}\}$. At $S^{++}$, both contacts are sliding
clockwise, and tangential forces are both positive. At $S^{--}$, both contacts
are sliding counter-clockwise, and tangential forces are negative. At $S^{+-}$,
contact 1 is sliding clockwise ($\tau_1 >0)$ and contact 2 is sliding
counter-clockwise ($\tau_2<0$). Finally, at $S^{-+}$, contact 1 is sliding
counter-clockwise ($\tau_1 < 0)$ and contact 2 is sliding clockwise
($\tau_2>0$).

For example, for configuration $S^{++}$, for which both tangential forces are
positive, the following equations of motion are obtained:
\begin{eqnarray}
  \label{eq:39}
  \mu m\ddot x = (\sin\alpha_1+ \mu\cos\alpha_1)\tau_1
  - (\sin\alpha_2 - \mu\cos\alpha_2)\tau_2\\
  \label{eq:40}
  \mu m\ddot y = (\cos\alpha_1 - \mu\sin\alpha_1)\tau_1
  + (\cos\alpha_2 + \mu\sin\alpha_2)\tau_2 - mg.
\end{eqnarray}
Similar equations are obtained for the other three sliding configurations.

Within the permanent sliding approximation, and since the duration of elastic
deformations is neglected, the dynamics of the system can be approximated as
series of transitions between these four sliding configurations. Transitions
between configurations occur each time the direction of sliding is reversed at a
contact. At the moment of this reversal, the velocity of the upper disk and the
support become equal. In practice, sliding stops momentarily, and a period of
elastic deformation of the skin begins. For the permanent-sliding approximation
to remain valid, this period of elastic deformation needs to be much shorter
than a typical duration of a sliding configuration. If support disks move
tangentially with a velocity much larger than the rotational velocity of the
upper disk this condition is met.

Assume the system, on average, stays at each sliding configuration
$S^{\pm\pm}$ during a time interval $T^{\pm\pm}$. The net angular displacement
that the upper disk undergoes in such interval can be obtain from the
average torque $\bar\T^{\pm\pm}$ acting on the disk during
$T^{\pm\pm}$.

Equations~\eqref{eq:37} and~\eqref{eq:38} can be used to estimate
$\bar\T^{\pm\pm}$. Unlike equations~\eqref{eq:13} and~\eqref{eq:14},
equations~\eqref{eq:37} and~\eqref{eq:38} have a well defined
translational-equilibrium solution (solutions for which $\dddot x =
\ddot y = 0$). In general, the equilibrium solution for the 3-disk
system is not unique, the solution involves solving a system of 3
equations of motion with four unknown forces
(see~\cite{mcnamara_indeterminacy_2005} for a more detailed
discussion). But, once sliding is assumed at both contacts,
equations~\eqref{eq:37} and~\eqref{eq:38} become a system of two
equations with two unknowns, from which equilibrium tangential forces
$\tau_1^{eq}$ and $\tau_2^{eq}$ can be obtained. For example, for
configuration $S^{++}$, equilibrium tangential forces can be obtained
by setting $\ddot x = \ddot y = 0$ in equations~\eqref{eq:39}
and~\eqref{eq:40} and solving the resulting system. Plugging the
solutions into equation~\eqref{eq:7} yields the equilibrium torque
\begin{equation}
  \label{eq:15}
  \T_{eq}^{++} = \frac{m g R \mu(\cos\tilt + \mu\sin\tilt)}{(1+\mu^2)\cos\theta_h},
\end{equation}
where $\tilt$ is the tilt angle, defined is Section~\ref{sec:methods} as
$\tilt = (\alpha_2-\alpha_1)/2$, and $\theta_h$ is half the aperture angle of
the contacts, defined as $\theta_h = (\alpha_2+\alpha_1)/2$. Similarly, the
equilibrium torques for configurations $S^{--}$, $S^{+-}$, and $S^{-+}$ are
obtained as
\begin{eqnarray}
  \label{eq:16}
  \T_{eq}^{--} = -\frac{m g R \mu(\cos\tilt - \mu\sin\tilt)}{(1+\mu^2)\cos\theta_h}\\
  \label{eq:17}
  \T_{eq}^{+-} = \frac{m g R \mu \sin\tilt}{\sin\theta_h + \mu\cos\theta_h}\\
  \label{eq:18}
  \T_{eq}^{-+} = -\frac{m g R \mu \sin\tilt}{\sin\theta_h - \mu\cos\theta_h}.
\end{eqnarray}
The equilibrium points just discussed are not true dynamical equilibrium points,
since the total torque on the upper disk is not zero. At these points, the
center of the upper disk is assumed in equilibrium, but there is angular
acceleration caused by the torque $\T_{eq}^{\pm\pm}$. We define these points to
be \emph{translational equilibrium} points, or TEPs. Each TEP is defined by the
point towards which the disk center is assumed to evolve at $S^{\pm\pm}$.  For
example, consider the case of support disks moving only tangentially to the
contact point (or only rotating). If the system stays at configuration
$S^{\pm\pm}$ long enough, the transient initial state will dissipate, and the
dynamics will converge to the TEP.\@ We are assuming here that fluctuations
around equilibrium will dissipate due to damping forces. In the more general
case, at which normal displacements are not restricted, fluctuations around the
TEP will not cease. Nevertheless, the TEP point will still be an attractor of
the dynamics. This statement can be justified by relaxing the constraints, and,
instead of dynamical equilibrium ($\ddot{x}=\ddot{y}=0$), only statistical
stationarity needs to be assumed. This is, we replace all quantities in
equations~(\ref{eq:37}) and~(\ref{eq:38}) by their mean values, and require that
the mean accelerations vanish, $\ddot{\bar x} = \ddot{\bar y} = 0$. In this
case, the TEP defines the mean values for contact forces around which
fluctuations take place.

Consider the following case to illustrate the nature of the TEP:\@
Assume supports are rotating clockwise, much faster than the upper
disk, with both contacts sliding at configuration $S^{++}$. After some
time, any oscillatory dynamical behavior dissipates due to damping,
the system arrives at the TEP, and the torque on the upper disk
remains constant at $\T_{eq}^{++}$, while the disk center remains
fixed. Since the disk suffers angular acceleration at this TEP, if
configuration $S^{++}$ were to be maintained much longer, the
rotational velocity of the upper disk would eventually catch up with
the velocity of the supports, after which the upper disk would perform
elastic rotational oscillations. What actually happens is that
supports are rapidly oscillating, and the system transitions to a new
sliding configuration before the velocity of the upper disk becomes of
the order of the typical velocity of the supports.

Although the system never actually reaches translational equilibrium,
the TEP torque can be used as an estimate for the mean torque, i.e.,
$\bar\T^{\pm\pm} \approx \T^{\pm\pm}_{eq}$. This amounts to
disregarding the cummulative contribution of torque fluctuations
around its TEP value. The duration $T^{\pm\pm}$ of each configuration
$S^{\pm\pm}$ is proportional to the correlation time of the motion
$t_c$.  This means that these estimates become increasingly accurate
as the correlation time $t_c$ increases. For very short correlation
times, the system has not enough time to reach the TEP, the forces are
practically random, and the disk does not rotate. Still, even if the
TEP is not reached, for medium values of $t_c$, torque values still
correlate with their values at the correspondign TEP $S^{\pm\pm}$, the
degree of the similarity improving the larger $t_c$ is.

Equations~\eqref{eq:15} through~\eqref{eq:18} depend explicitly on $\tilt$. The
torques $\T^{\pm\pm}_{eq}$ are the analogue to the friction limits $F_1$ and $F_2$ of
the EPO.\@ When the system is tilted, the Amonton limits for different sliding
configuration become asymmetric, introducing a sliding bias. Under the permanent
sliding approximation, the total mean torque on the system can be calculated as
\begin{equation}
  \label{eq:11}
  \bar\T = \frac{ \bar\T^{++} T^{++}
    + \bar\T^{++} T^{--} + \bar\T^{+-} T^{+-} + \bar\T^{-+} T^{-+} }{T^{++}+T^{--}+T^{+-}+T^{-+}},
\end{equation}
where the $\bar\T^{\pm\pm}$ are given by equations~\eqref{eq:15}
through~\eqref{eq:18}, and the $T^{\pm\pm}$ are the mean duration of
each sliding configuration $S^{\pm\pm}$. Assuming the system reaches a
steady state, with a stationary mean rotational velocity for the upper
disk, we require that the total mean torque $\bar\T$ vanishes, i.e.,
\begin{equation}
  \label{eq:41}
  \bar\T = 0.
\end{equation}
Since the mean torques $\bar\T^{\pm\pm}$ are given by their TEP
values, this condition imposes a constraint on the mean times
$T^{\pm\pm}$.

At each $S^{\pm\pm}$, the upper disk accelerates under the effect of the
torque. Thus, times $T^{\pm\pm}$ and the rotational velocity of the upper disk
are related. The details of this relation are explained in
Sections~\ref{sec:synchr-rotat-supp} and~\ref{sec:asynchr-rotat-supp} through
the introduction of two example cases. But, without knowing such details,
predictions about the sign of $\mrvel$ using can still be made using simple
arguments.

Consider tilting the system slightly from $\tilt = 0$. When there is
no tilt, $\bar\T^{++}_{eq} = \bar\T^{--}_{eq}$ and
$\bar\T^{+-}_{eq}=\bar\T^{-+}_{eq} = 0$, per equations~\eqref{eq:15}
through~\eqref{eq:18}. For randomly vibrating supports it is expected,
from symmetry arguments, that at zero tilt all times $T^{\pm\pm}$ are
equal, because the mean total torque $\bar\T$ and the mean rotational
velocity of the upper disk are both zero. As the system is slightly
tilted, all times $T^{\pm\pm}$ can be assumed to remain initially
unchanged. When $\T^{\pm\pm}_{eq}$ are evaluated at the new angle
$\theta_T$, using equation~\eqref{eq:11}, the net torque on the upper
disk is calculated as
\begin{equation}
  \label{eq:19}
  \eqalign{
    \bar\T^* &= \frac{1}{4}\left( \T^{++} + \T^{--} + \T^{+-} + \T^{-+}  \right) \\
    &= \frac{mgR\mu^2(\mu^2+(1+\mu^2)\cos 2\theta_h)\sec^3\theta_h\sin\tilt
    }
    {2(1+\mu^2)(\mu^2-\tan^2\theta_h)}.}
\end{equation}
A nonzero mean torque $\bar\T^*$ then acts on the upper disk,
immediately after tilting the system. The rotational acceleration then
becomes nonzero under the effect of $\bar\T^*$, until the times
$T^{\pm\pm}$ adjust to comply with condition~\eqref{eq:41}. For the
parameter values used in the simulations of
Section~\ref{sec:numer-simul}, the slight-tilt torque $\bar\T^*$ is
negative, resulting in a negative rotational velocity, as presented in
that same section.

Another illustrative situation worth considering is that of constant
normal forces. If the centers of all disks are fixed, and the supports
are only allowed to rotate, normal forces at the contacts remain
constant. Assume furthermore that $n_1\neq n_2$, so that reflection
symmetry is broken. Despite this asymmetry, the Amonton limit for the
torque remains fixed at $\mu(n_1 + n_2)$ for all sliding
configurations. This is similar to the symmetric EPO with $\abs{F_1} =
\abs{F_2}$, and there exists no sliding bias. It was numerically
verified that, in such case (when disk translations are forbidden),
the upper disks never accumulates rotations. This is due to the fact
that sliding limits only become asymmetric when contact forces evolve
toward their TEP values.

It has been stated that times $T^{\pm\pm}$ adjust in order to satisfy
the condition~\eqref{eq:41}, but a description of the mechanism of
this adjustment has yet to be provided. In the next sub-sections, two
special cases are presented for which an approximate analytical
solution for the rotational velocity of the upper disk can be found:
1) the supports rotate harmonically in phase, and 2) the supports
rotate harmonically, with opposite phases. It is found that the upper
disk rotates in opposite directions in each of these cases, indicating
that the sense and velocity of rotation depends strongly of the
details of the excitation. As will be discussed, this is due to the
fact that the times $T^{\pm\pm}$ depend on the relative phase between
the rotation of the support disks. Although the vibration of the
supports in these two cases is not random, the mechanism of adjustment
for the $T^{\pm\pm}$ is similar as for the random case. The advantage
of considering harmonic excitation is that the system becomes solvable
using straightforward calculations.
\subsection{Synchronous rotation of the supports}\label{sec:synchr-rotat-supp}
\begin{figure}
  \centering
  \includegraphics[width=\scale\linewidth]{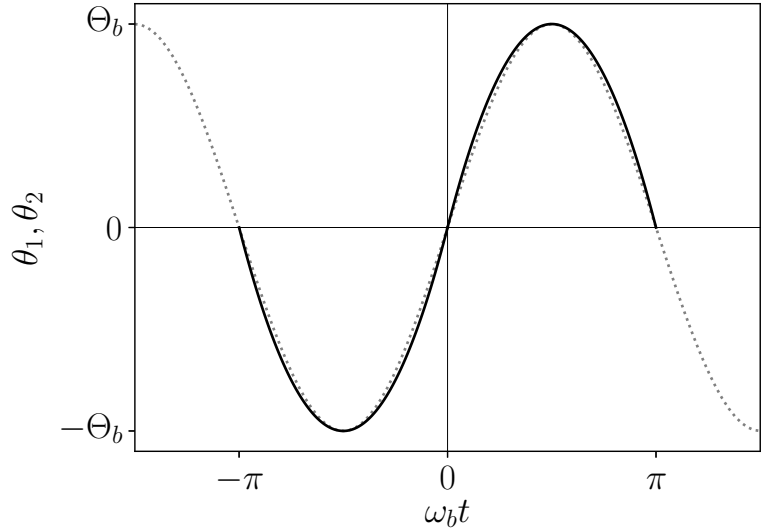}
  \caption{The rotational displacement of the support disks (dotted line) and
    the piecewise-parabolic approximation to the motion (solid
    line.)}\label{fig:sine-parabola-approx}
\end{figure}
Consider, first, the case where the centers of the support disks are
fixed, and they rotate periodically in phase. The upper disk is
allowed to translate and rotate. The angular excursions of the support
disks are given by the equation
\begin{equation}
  \label{eq:20}
  \theta_1 = \theta_2 = \Theta_b\sin(\omega_b t),
\end{equation}
where $\theta_1$ is the angle of support 1, $\theta_2$ is the angle of
support 2, $\Theta_b$ is the amplitude of angular oscillations, and
$\omega_b$ is the angular frequency. The period of the oscillations is
$T = 2\pi/\omega_b$. In order to simplify calculations,
equation~\eqref{eq:20} can be piecewise approximated by parabolas, as
shown in Figure~\ref{fig:sine-parabola-approx}. The fitted parabolas
are required to match the minimum and maximum of the sine function, as
well as the crossings at the $t$-axis. In the interval from $-T/2$ to
$T/2$, equation~\eqref{eq:20} is, then, approximated by the parabolas
\begin{equation}
  \label{eq:21}
  \theta_1 =
  \cases{
  \frac{4\Theta_b\omega_b^2}{\pi^2}t^2 + \frac{4\Theta_b\omega_b}{\pi}t &
    for $-\pi<\omega_b t < 0$ \\
    \frac{-4\Theta_b\omega_b^2}{\pi^2}t^2 + \frac{4\Theta_b\omega_b}{\pi}t &
    for $0 < \omega_b t < \pi$.
}
\end{equation}
The velocity at the contact point can be found by differentiating
equation~\eqref{eq:21},
\begin{equation}
  \label{eq:22}
  v_1 = - R\frac{d\theta_1}{dt} =
  \cases{
    -\frac{8R\Theta_b\omega_b^2}{\pi^2}t - \frac{4R\Theta_b\omega_b}{\pi} &
    for $-\pi<\omega_b t < 0$\\
    \frac{8R\Theta_b\omega_b^2}{\pi^2}t - \frac{4R\Theta_b\omega_b}{\pi} &
    for $0<\omega_b t < \pi$,
  }
\end{equation}
where $R$ is the disk radius. Equation~\eqref{eq:22} describes a triangle
wave. This particular approximation was chosen because it allows for an easy
analytic solution for the velocity of the upper disk.
\begin{figure}
  \centering
  \includegraphics[width=\scale\linewidth]{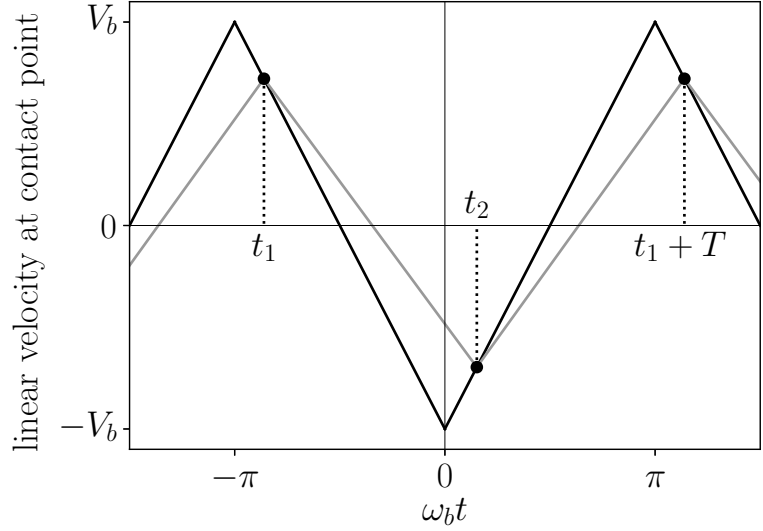}
  \caption{The velocity of the support disks (black) and the upper disk (grey)
    as they evolve in time.}\label{fig:sync}
\end{figure}
For supports rotating in phase, contacts can never slide in opposite
directions. This immediately excludes the possibility of reaching sliding
configurations $S^{+-}$ and $S^{-+}$, thus $T^{+-}$ and $T^{-+}$ are set to zero
accordingly. Notice that this is an important difference with the case of
randomly vibrating supports, and shows that the times $T^{\pm\pm}$ depend strongly on
the nature of the vibration.

Regardless of the initial condition, the system always reaches a
periodic steady state, that has the same period as that of the
driving. Figure~\ref{fig:sync} shows the typical stationary-state
behavior of the velocities for all disks during a period of
oscillation $T$. The times labeled as $t_1$ and $t_2$ are those at
which the velocity of the upper disk equals the velocity of the
supports, identified by the intersection of the black (supports
velocity) and gray lines (velocity of the upper disk). Under the
assumption of permanent sliding, these crossing points mark the
transitions between the two allowed sliding configurations. When the
velocities of the upper disk and the supports become equal, the
relative tangential motion reverses, and a transition takes place.

Before $t_1$, the supports move faster than the upper disk and rotate
ahead of it. Tangential forces are positive, and the sliding
configuration is $S^{++}$. After $t_1$, and up to $t_2$, the supports
move slower than the upper disk, thus making the tangential forces
negative. The corresponding sliding configuration is $S^{--}$ in this
case. At $t_2$, the sliding reverses again, and the configuration
transitions back to $S^{++}$. At $t_1 + T$, a cycle is completed and
the systems returns to the same state as at $t_1$.

While at configuration $S^{--}$, between $t_1$ and $t_2$, there exists
a torque that acts on the uppers disk. This torque is assumed constant
and equal to $\T^{--}_{eq}$, given by~\eqref{eq:16}. Configuration
$S^{--}$ lasts during a total time $T^{--} = t_2 - t_1$, at the end of
which the velocity of the upper disk has suffered a net change of
$R\T^{--} T^{--}/I$, where $I$ is the moment of inertia. Since at
$t_1$ and $t_2$ the velocities of the support and the upper disk must
match, the velocity at both transition points is related by the
equation
\begin{equation}
  \label{eq:23}
  v_1(t_1) + R\T^{--} T^{--}/I = v_1(t_2),
\end{equation}
where $v(t)$ is the function defined in equation~\eqref{eq:22}.
Equation~\eqref{eq:23} states that the velocity of the upper disk at
$t_1$, given by $v_1(t_1)$, plus the net change in velocity, must
match the velocity of the supports at $t_2$, given by
$v_1(t_2)$. Equations of this type are referred here to as
velocity-matching equations.

From $t_2$, and up to $t_1 + T$, the system is at configuration $S^{++}$. The
net velocity change suffered during the interval $T^{++} = t_1 + T - t_2$ is
given by $R\T^{++} T^{++}/I$, where $\T^{++}$ is the constant torque
(given by equation~\eqref{eq:15}) and $I$ is again the moment of inertia. The
velocity-matching equation between times $t_2$ and $t_1 + T$ is
\begin{equation}
  \label{eq:24}
  v_1(t_2) + R\T^{++} T^{++}/I = v_1(t_1 + T) = v_1(t_1),
\end{equation}
where the last equality comes from the periodicity of the motion.

Solving equations~\eqref{eq:23} and~\eqref{eq:24}, for $t_1$ and
$t_2$, yields
\begin{eqnarray}
  \label{eq:25}
  t_1 = \frac{\pi \T^{++} (\pi^2 \T^{--} -
  4mR^2\Theta_b\omega_b^2)}{4mR^2\Theta_b(\T^{++} - \T^{--})\omega_b^3}\\
  \label{eq:43}
  t_2 = \frac{\pi \T^{++} (\pi^2 \T^{--} + 4mR^2\Theta_b\omega_b^2)}{4mR^2\Theta_b(\T^{++} - \T^{--})\omega_b^3}.
\end{eqnarray}
From which $T^{++}$ and $T^{--}$ can be calculated as
\begin{eqnarray}
  \label{eq:26}
  T^{++} =  t_1 + T - t_2 = -\frac{\T^{--}}{\T^{++} - \T^{--}}T\\
  \label{eq:27}
  T^{--} = t_2 - t_1 = \frac{\T^{++}}{\T^{++} - \T^{--}}T.
\end{eqnarray}
With $t_1$ and $t_2$ given by equations~\eqref{eq:25}
and~\eqref{eq:43}, and under the assumption of constant torques, the
instantaneous rotational velocity of the upper disk can be written as
\begin{equation}
  \label{eq:28}
  \rvel(t) =
  \cases{
    \frac{v_1(t_1)}{R} + \frac{\T^{--}(t - t_1)}{I} & for $t_1 < t < t_2$\\
    \frac{v_1(t_2)}{R} + \frac{\T^{++}(t - t_2)}{I} & for $t_2 < t < t_1 + T$.
  }
\end{equation}
Equation~\eqref{eq:28} yields the correct transition values
$\rvel(t_1) = v_1(t_1)/R$ and $\rvel(t_2) = v_1(t_2)/R$.  The mean
rotational velocity of the upper disk is found by integrating
equation~\eqref{eq:28} over a complete period,
\begin{equation}
  \label{eq:29}
  \mrvel = \frac{1}{T}\int_{t_1}^{t_1 + T}\rvel(t)dt
  = \frac{4 \Theta_b\omega_b}{\pi}\frac{\T^{++} + \T^{--}}{\T^{++} - \T^{--}}.
\end{equation}
Integrating, and using equations~\eqref{eq:15} and~\eqref{eq:16}, results in
\begin{equation}
  \label{eq:30}
  \mrvel = \frac{4\mu \Theta_b\omega_b \tan\tilt}{\pi}.
\end{equation}

\begin{figure}
  \centering
  \includegraphics[width=\scale\linewidth]{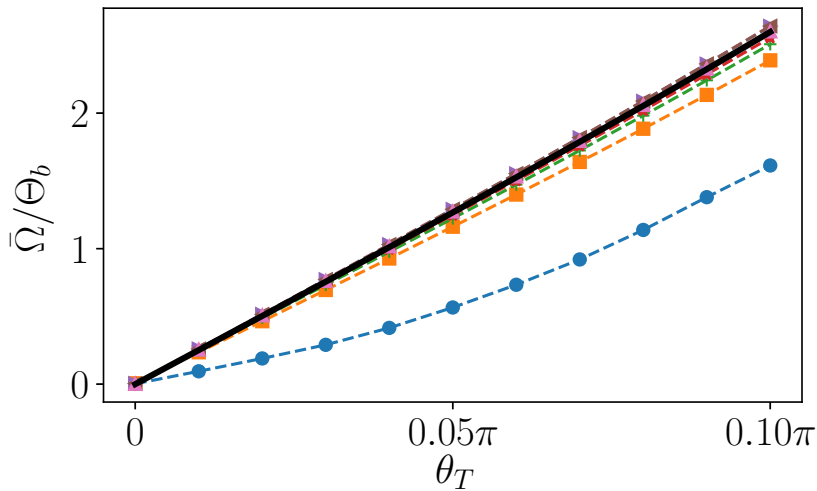}
  \caption{The rotational velocity of the upper disk $\mrvel$, normalized by the
    oscillation amplitude of the supports $\Theta_b$, versus the system tilt
    $\tilt$. Numerical simulations, with the supports rotating completely in
    phase are show (dashed lines) for different values of the oscillating
    amplitude $\Theta_b$: $\Theta_b=1\sci{-3}1$ (circles), $\Theta_b=4\sci{-3}$
    (squares), $\Theta_b=7\sci{-3}$ (pluses), $\Theta_b=1\sci{-2}$ (stars),
    $\Theta_b=4\sci{-2}$ (triangles pointing right), $\Theta_b=7\sci{-2}$ (triangles
    pointing left), and $\Theta_b=1\sci{-1}$ (triangles pointing down). The
    agreement between simulations and the predicted velocity given by
    equations~\eqref{eq:30} (solid line) is excellent.}\label{fig:sync-vel}
\end{figure}
Figure~\ref{fig:sync-vel} shows a comparison between the velocity
predicted by equation~\eqref{eq:29} and results obtained from
numerical simulations under the appropriate excitation conditions.
Velocities obtained from numerical simulations approach the value
predicted by equation~\eqref{eq:29} as the amplitude of excitation
$\Theta_b$ increases. The agreement for large $\Theta_b$ becomes
better since the assumption of permanent sliding is valid for large
excitation intensities. For low amplitudes, the contacts spend
non-negligible times in the elastic phase, and the approximation of
constant sliding breaks down.

Notice that the rotational velocity in the case of synchronous
rotation of the supports is always positive, and, thus, opposite in
sign to case of random vibration presented in
Section~\ref{sec:numer-simul}. It is possible to apply the same
small-tilt analysis used to predict the sign of the velocity under
random vibration to the present case of in-phase rotation of the
supports. As mentioned before, for supports rotating in phase, sliding
configurations $S^{+-}$ and $S^{-+}$ are inaccessible to the
system. The instantaneous mean torque starting from zero tilt and
slightly tilting the system is now calculated as
\begin{equation}
  \label{eq:31}
  \bar\T^* = \frac{\T^{++} + \T^{--}}{2} = \frac{mgR\mu \sin\tilt}{(1+\mu^2)\cos\theta_h},
\end{equation}
which, for the values of the parameters employed in the simulations,
is now positive. Upon tilting the system, a positive torque acts on
the upper disk, increasing its mean rotational velocity, until times
$T^{++}$ and $T^{--}$ adjust, and the dynamical behavior becomes
periodic. The values of $T^{++}$ and $T^{--}$ in the periodic
stationary state are given by equations~\eqref{eq:26}
and~\eqref{eq:27}. It is easy to verify that these values comply with
condition~\eqref{eq:41}, and make the mean torque on the upper disk
zero.

A geometric interpretation of how the system converges to periodic
behavior can be given by analyzing Fig.~\ref{fig:sync}. Let us assume
that the system starts in a state where $T^{++} = T^{--}$, but $\tilt
\neq 0$. The torque on the upper disk is given by
equation~\eqref{eq:31}, and is positive. When $\mrvel$ grows under the
effect of the torque, the lines describing the velocity of the upper
disk in Figure~\ref{fig:sync} will shift up. This upward shift moves
the points ($t_1$, $\rvel_1(t_1)$) and ($t_2$, $\rvel_2(t_2)$) towards
the maximum of the triangle wave. This decreases the time interval
$T^{++}$, while simultaneously increasing $T^{--}$, effectively
decreasing the total torque on the upper disk. The process will
continue until $T^{++}$ and $T^{--}$ reach their stationary values, at
which the mean torque over a period vanishes.
\subsection{Asynchronous rotation of the supports}\label{sec:asynchr-rotat-supp}
\begin{figure}
  \centering
  \includegraphics[width=\scale\linewidth]{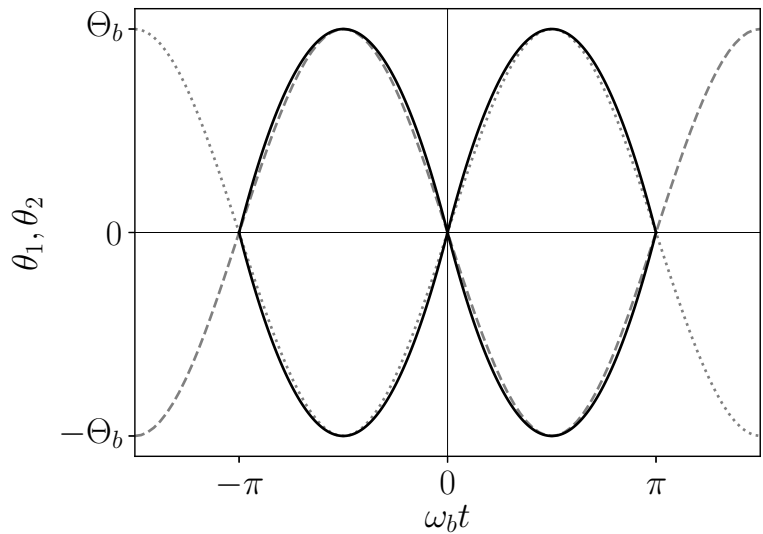}
  \caption{The rotational displacement of the support disks (dotted lines) and
    the piecewise-parabolic approximations to the motion (solid
    line.)}\label{fig:sine-parabola-approx-async}
\end{figure}
The same calculations used in Section~\ref{sec:synchr-rotat-supp} can
be employed in the case of support disks rotating completely out of
phase. Now the supports rotate in opposite directions, i.e., $\theta_2
= -\theta_1$, and their angular velocities are also opposite in sign,
$v_2 = -v_1$. In the piece-wise parabolic approximation, $\theta_1$
and $v_1$ are still given by equations~\eqref{eq:21}
and~\eqref{eq:22}, respectively. This case is illustrated in
Figure~\ref{fig:sine-parabola-approx-async}, where the angles of both
supports are shown, together with the corresponding piece-wise
parabolic approximation for their sinusoidal motion.
\begin{figure}
  \centering
  \includegraphics[width=\scale\linewidth]{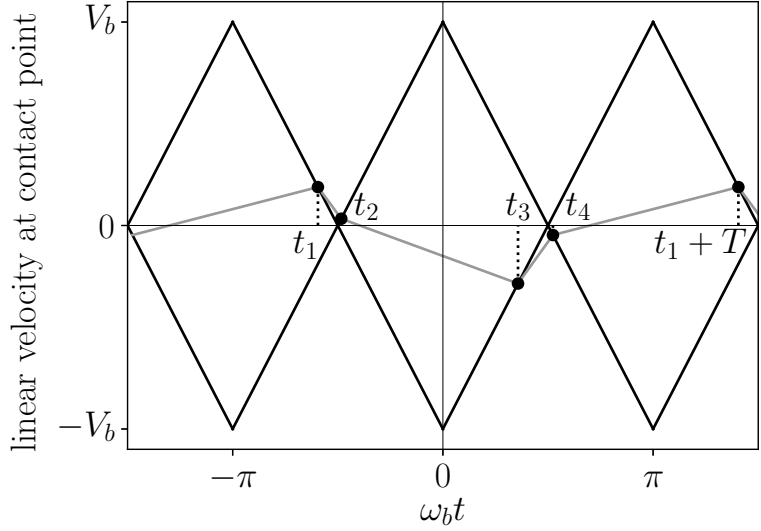}
  \caption{The velocity of the support disks (black) and the velocity of the
    upper disk (gray) as they evolve in time.}\label{fig:async}
\end{figure}
Figure~\ref{fig:async} shows the velocities of the supports and the
velocity of the upper disk in a cycle of period $T$, once the periodic
state has been reached. Since the supports are not in phase,
configurations $S^{+-}$ and $S^{-+}$ are now reachable by the
system. Referring to Figure~\ref{fig:async}, and starting at $t_1$,
the angular velocity of the upper disk is larger than that of both
supports. Tangential forces are then negative, and the system is at
configuration $S^{--}$. At $t_2$, the velocity of the upper disk
becomes equal to that of support 2. The tangential force of contact 1
remains negative, but the sliding direction at contact 2 reverses, and
the system transitions into configuration $S^{-+}$. At $t_3$, the
velocity of the upper disk equals the velocity of support disk 1,
reversing the sliding at contact 1, making both tangential forces
positive. The system then transitions onto configuration $S^{++}$. At
$t_4$, the velocity of the upper disk equals that of support 2 again,
making tangential force at contact 2 negative, while tangential force
at contact 1 remains positive. The system then transitions onto
configuration $S^{+-}$. At $t_1 + T$, the cycle ends and the system
returns to configuration $S^{--}$.  The velocity-matching equations
introduced in Section~\ref{sec:synchr-rotat-supp} now take the form
\begin{eqnarray}
  \label{eq:32}
  v_1(t_1) + R\T^{--} T^{--}/I = v_2(t_2)\\
  v_2(t_2) + R\T^{-+} T^{-+}/I = v_1(t_3)\\
  v_1(t_3) + R\T^{++} T^{++}/I = v_2(t_4)\\
  \label{eq:33}
  v_2(t_4) + R\T^{+-} T^{+-}/I = v_1(t_1+T) = v_1(t_1),
\end{eqnarray}
where $T^{--}=t_2-t_1$, $T^{-+}=t_3-t_2$, $T^{++}=t_4-t_3$, and $T^{+-}=t_1+T-t_4$. The
velocity $v_2$ is obtained from the condition $v_2 = -v_1$, with $v_1$ given by
equation~\eqref{eq:22}.

The angular velocity of the upper disk is a piece-wise continuous
function with four continuous intervals, and can be written as
\begin{equation}
  \label{eq:34}
  \rvel(t) =
  \cases{
    \frac{v_1(t_1)}{R} + \frac{\T^{--}(t - t_1)}{I} & for $t_1 < t < t_2$\\
    \frac{v_2(t_2)}{R} + \frac{\T^{-+}(t - t_2)}{I} & for $t_2 < t < t_3$\\
    \frac{v_1(t_3)}{R} + \frac{\T^{++}(t - t_3)}{I} & for $t_3 < t < t_4$\\
    \frac{v_2(t_4)}{R} + \frac{\T^{+-}(t - t_4)}{I} & for $t_4 < t < t_1 + T$.
  }
\end{equation}
The torques acting during each sliding configuration are given by
equations~\eqref{eq:15} through~~\eqref{eq:19}, and are assumed
constant, under the approximation of permanent sliding.

Following the same process discussed in
Section~\ref{sec:synchr-rotat-supp}, the transition times $t_1$
through $t_4$ can be obtained by solving equations~\eqref{eq:32}
through~\eqref{eq:33}. Using these solutions together with
equation~\eqref{eq:34} for $\rvel$, the integral~\eqref{eq:29} can be
done to obtain an expression for $\mrvel$. After integrating and
expanding to first order in $\tilt$, the approximate mean rotational
velocity takes the form
\begin{equation}
  \label{eq:35}
  \rvel\approx\frac{8 \mu \left(\mu ^2+1\right) \Theta _b \theta _T
    \cos ^2\left(\theta _h\right)}{\left(\mu ^2+1\right) \cos \left(2
      \theta _h\right)+\mu ^2-1}.
\end{equation}

\begin{figure}
  \centering
  \includegraphics[width=\scale\linewidth]{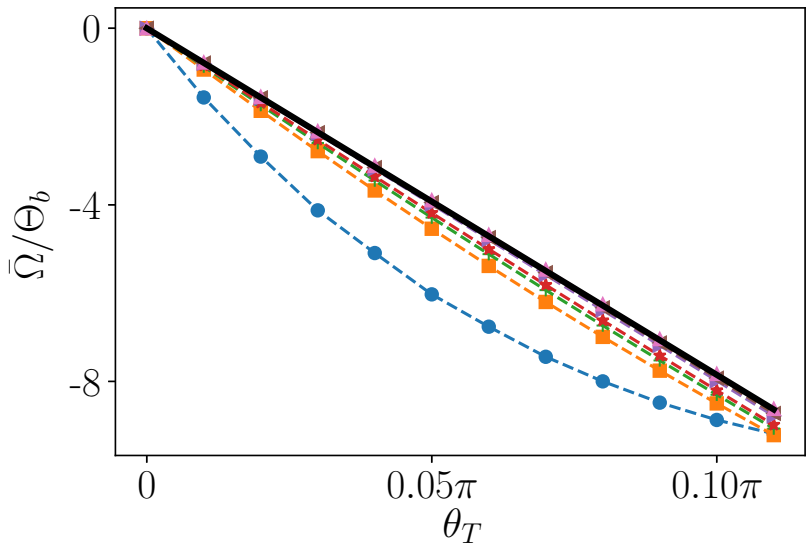}
  \caption{The rotational velocity of the upper disk $\mrvel$,
    normalized by the oscillation amplitude of the supports
    $\Theta_b$, versus the system tilt $\tilt$. Numerical simulations,
    with the supports rotating completely out of phase, are show
    (dashed lines) for different values of the oscillating amplitude
    $\Theta_b$: $\Theta_b=1\sci{-3}$ (circles), $\Theta_b=4\sci{-3}$
    (squares), $\Theta_b=7\sci{-3}$ (pluses), $\Theta_b=1\sci{-2}$
    (stars), $\Theta_b=4\sci{-2}$ (triangles pointing right),
    $\Theta_b=7\sci{-2}$ (triangles pointing left), and
    $\Theta_b=1\sci{-1}$ (triangles pointing down). The agreement
    between simulations and the predicted velocity given by
    equations~\eqref{eq:35} (solid line) is
    excellent.}\label{fig:async-vel}
\end{figure}
Figure~\ref{fig:async-vel} shows a comparison between the velocity
predicted by equation~\eqref{eq:35} and results obtained from
numerical simulations. Velocities obtained from numerical simulations
approach the value predicted by equation~\eqref{eq:35} as $\Theta_b$
increases and the permanent-sliding approximation becomes increasingly
accurate.  The agreement for large amplitudes $\Theta_b$ is, again,
excellent.

Contrary to the case discussed in Section~\ref{sec:synchr-rotat-supp},
when the supports rotate out of phase the rotational velocity is
always negative. For this reason, out-of-phase excitation resembles
the case of random vibration presented in
Section~\ref{sec:numer-simul}. In both the random and the out of phase
cases, the system can access all four sliding configurations
$S^{\pm\pm}$. The analysis to predict the sign of the velocity, based
on equation~\eqref{eq:19}, is the same for both cases, and predicts a
negative velocity in both of them.
\section{Conclusion}\label{sec:conclusion}
In this work, a new rotational ratcheting mechanism was reported,
which occurs in a simple packing consisting of a single disk supported
against gravity by other two. It was shown, using numerical
simulations that, if the supports are vibrated and the system is
tilted, the upper disk acquires a non-zero mean rotational velocity,
even though the vibration is completely left-right symmetric. It was
also shown that the details of the vibration are important, as
changing them may lead to velocity inversion of the upper disk.

Similarly to the case of the elasto-plastic oscillator (EPO), for
rotation to appear, friction forces must be asymmetric for different
directions of sliding. This asymmetry originates in the correlation
between contact forces and the direction at which contacts slide. The
notion of translational equilibrium points (TEP) was introduced to
explain the origin of these correlations. The TEPs define
translational equilibrium points towards which the system converges
during sliding. Since each of the two contacts can slide in two
directions, there exist four different sliding configurations, each
one with a different TEP. In this sense, the 3-disk system may be
regarded as a generalization of the EPO, with four possible sliding
configurations instead of two.

One might attempt to apply to this 3-disk system, some of the
approaches that have been previously used to analyze the EPO under
random loading (see for
example~\cite{karnopp_plastic_1966,bouc_ratcheting_1998,ditlevsen_plastic_1993}). These
techniques, however, assume rare visits to the plastic domain, a limit
case completely opposite to the one elaborated upon in this
work. Rotation in the 3-disk system requires contacts to be saturated
(sliding) a sizeable fraction of the time, since it is only during
sliding that correlations among forces appear, and the system evolves
towards the TEP.\@ This makes such mentioned techniques hard to adapt
to the conditions of vibration described here.

Here, two simple deterministic cases were solved under the assumption
of constant sliding. More work is required to obtain an expression for
the rotational velocity with supports that are vibrated randomly. This
case will be addressed in future work.

As a final note, we have recently observed a related phenomena of self-organized
rotations in disk packings (manuscript in preparation). We found that when
two-dimensional disk packings are vibrated from the bottom, each of the disks
within the packing acquires a rotational velocity that depends on the local
configuration of contacts of each disk. Given that the contact network of a
packing is essentially random, rotations in such packings are an example of
noise rectification in disordered systems. Since large packings are
generalizations of the simple packing presented here, one can expect that some
of the mechanisms described in this work remain at play for packings of more
than three disks.


\ack

GPM was supported by a PhD fellowship from CONACYT M\'exico. We acknowledge the
use of computational resources on clusters ``Xiuhcoatl'' and ``Kukulc\'an'' of
CINVESTAV.

\section*{References}
\bibliographystyle{iopart-num}
\bibliography{bibliography}

\end{document}